\documentclass[conference]{IEEEtran}

\usepackage{graphicx}
\usepackage{amsmath,amssymb}
\usepackage{siunitx}
\usepackage{upgreek}
\usepackage[noend]{algpseudocode}
\usepackage{algorithm}
\usepackage{multirow}
\usepackage{url}
\usepackage{balance}
\let\labelindent\relax
\usepackage{enumitem}
\usepackage{xcolor}
\usepackage{soul}
\usepackage{hyperref}
\usepackage{caption}
\usepackage{subcaption}
\usepackage{booktabs}
\usepackage{makecell}
\usepackage[colorinlistoftodos,prependcaption,textsize=small]{todonotes}

\ifCLASSINFOpdf
\else
\fi

\begin{document}
%

\title{Neutrino Oscillation Parameter Estimation \\Using Structured Hierarchical Transformers}

\author{\IEEEauthorblockN{Giorgio Morales$^{1}$, Gregory Lehaut$^{2}$, Antonin Vacheret$^{2}$, Frédéric Jurie$^{1}$, and Jalal Fadili$^{1}$}
\IEEEauthorblockA{$^{1}$ \quad Normandie Univ, UNICAEN, ENSICAEN, CNRS, GREYC. Caen, France \\
$^{2}$ \quad Normandie Univ, ENSICAEN, UNICAEN, CNRS/IN2P3, LPC Caen. Caen, France}
}


\maketitle

\begin{abstract}

Neutrino oscillations encode fundamental information about neutrino masses and mixing parameters, offering a unique window into physics beyond the Standard Model.
Estimating these parameters from oscillation probability maps is, however, computationally challenging due to the maps’ high dimensionality and nonlinear dependence on the underlying physics.
Traditional inference methods, such as likelihood-based or Monte Carlo sampling approaches, require extensive simulations to explore the parameter space, creating major bottlenecks for large-scale analyses. 
In this work, we introduce a data-driven framework that reformulates atmospheric neutrino oscillation parameter inference as a supervised regression task over structured oscillation maps. 
We propose a hierarchical transformer architecture that explicitly models the two-dimensional structure of these maps, capturing angular dependencies at fixed energies and global correlations across the energy spectrum. 
To improve physical consistency, the model is trained using a surrogate simulation constraint that enforces agreement between the predicted parameters and the reconstructed oscillation patterns. 
Furthermore, we introduce a neural network-based uncertainty quantification mechanism that produces distribution-free prediction intervals with formal coverage guarantees. 
Experiments on simulated oscillation maps under Earth-matter conditions demonstrate that the proposed method is comparable to a Markov Chain Monte Carlo baseline in estimation accuracy, with substantial improvements in computational cost (around 240$\times$ fewer FLOPs and 33$\times$ faster in average processing time). 
Moreover, the conformally calibrated prediction intervals remain narrow while achieving the target nominal coverage of 90\%, confirming both the reliability and efficiency of our approach.
\footnote{This paper is a preprint (accepted to appear in the International Joint Conference on Neural Networks 2026). IEEE copyright notice. Personal use of this material is permitted. Permission from IEEE must be obtained.}
\end{abstract}

\begin{IEEEkeywords}
Neutrino oscillation maps, Transformer Networks, Uncertainty Quantification 
\end{IEEEkeywords}


\section{Introduction} \label{sec:intro}

Neutrino ($\nu$) oscillations provide one of the most compelling pieces of evidence for physics beyond the Standard Model, demonstrating that neutrinos are massive and that their flavors mix as they propagate~\cite{Pontecorvo1968,Ahmad2002}. 
Determining the values of the oscillation parameters that govern this mixing is a central goal of contemporary neutrino physics, 
with precise measurements potentially informing the ordering of neutrino masses, the existence of charge-parity (CP) violation in the lepton sector, and the possible presence of new physics~\cite{Esteban2020,deSalas2021,PDG2024}.

Experimental efforts over the past decades have made remarkable progress, yet challenges remain~\cite{An2016, Abe2018, Acero2022}. 
The inference of oscillation parameters from experimental data is inherently difficult because the measured signals depend non-trivially on both neutrino energy and trajectory through the Earth~\cite{Nunokawa2008,Gonzalez-Garcia2008,Qian2012}. 
To capture this dependence, neutrino experiments often represent oscillation probabilities as structured maps that summarize the flavor transition probabilities across a range of energies and propagation angles~\cite{Rott2015}. 
These maps are rich in information but also highly complex, making parameter estimation a non-trivial task.

Traditionally, parameter inference relies on likelihood analyses that compare observed data with theoretical predictions from large-scale Monte Carlo (MC) simulations~\cite{Esteban2017,Acero2024}. 
While robust, these approaches are computationally expensive and require significant resources to explore high-dimensional spaces. 
This creates bottlenecks for large experiments, especially when real-time analysis or repeated hypothesis testing is required.
Recent observations of ultra-high-energy cosmic neutrinos by KM3NeT~\cite{KM3NeT_Collaboration2025-ca} further highlight the need for efficient inference methods that can handle broad energy ranges and deliver reliable estimates.
To address these limitations, machine learning-based approaches have been increasingly explored in neutrino physics, with applications ranging from event reconstruction and classification~\cite{Peterson:2023ayg,Heyer:2024nkq} to the direct estimation of oscillation parameters~\cite{Abbasi2025,Gavrikov:2025rps}. 

In this work, we propose a data-driven method that poses atmospheric neutrino oscillation parameter inference as a supervised regression task from oscillation probability maps. 
By leveraging simulation-based training, we enable the model to learn mappings between oscillation patterns and underlying physical parameters without the need for explicit likelihood evaluation. 
In particular, we develop a hierarchical transformer architecture that exploits the structured nature of oscillation maps; i.e., each energy–angle slice carries distinct physical meaning, and the flavor channels encode correlated but non-redundant transition probabilities. 
Our approach is designed to capture both local dependencies within each flavor transition channel and global interactions across channels, ultimately producing robust estimates of the oscillation parameters.

\begin{figure*}[t]
    \centering
    \includegraphics[width=.8\linewidth]{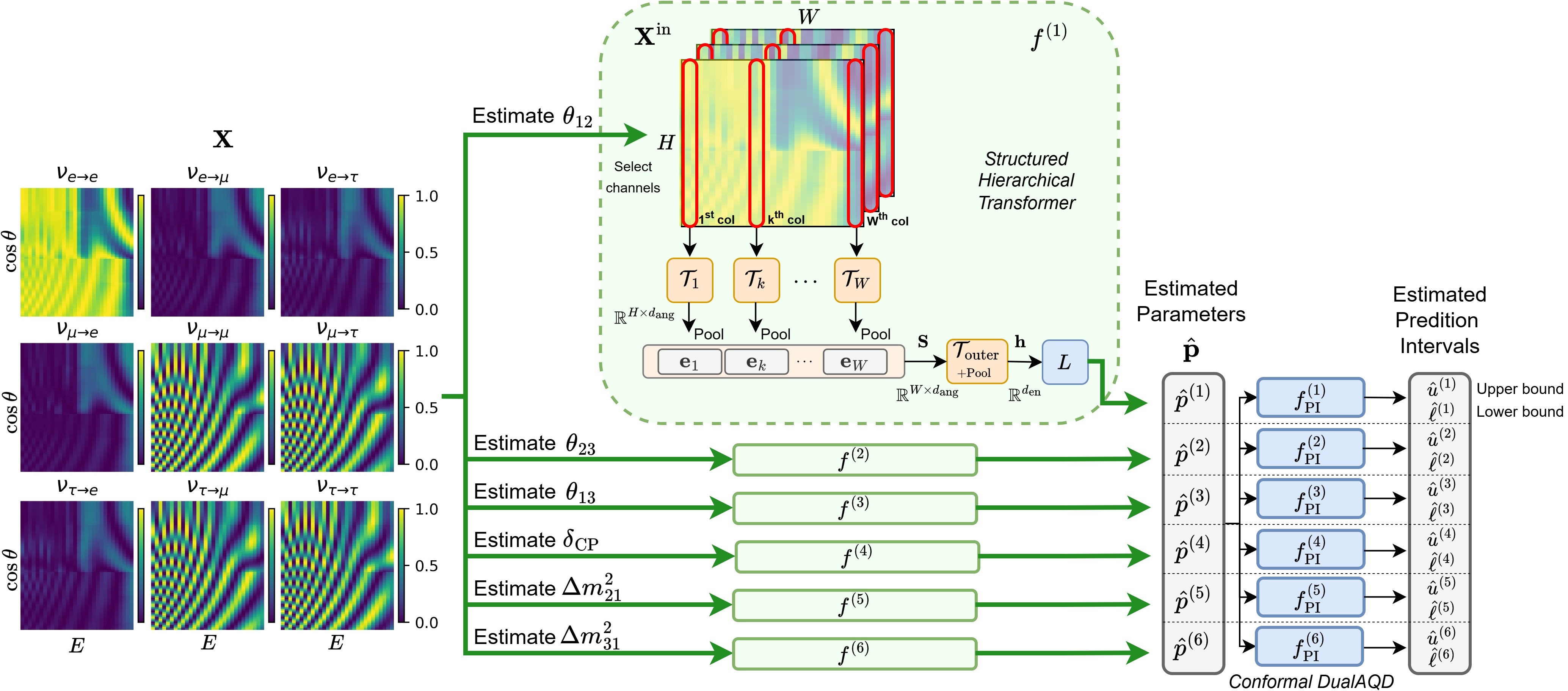}
    \vspace{-1ex}
    \caption{Overview of the proposed $\nu$ oscillation parameters estimation}
    \label{fig:overview}
    \vspace{-1ex}
\end{figure*}

To ensure stable and physically consistent learning, the training process is assisted by differentiable surrogate simulators that approximate oscillation probabilities with high fidelity, allowing for end-to-end gradient-based optimization.
In addition, we integrate an uncertainty quantification scheme based on a prediction interval-generation neural network and conformal prediction, which provides statistically calibrated prediction intervals around each estimated oscillation parameter. 
This is particularly important in oscillation parameter inference, where reliable uncertainty estimates are critical for downstream physics analyses and for assessing the robustness of the inferred parameters. 
Once trained, inference with the proposed framework is orders of magnitude faster than traditional likelihood-based analyses, enabling practical parameter estimation.
Our results on simulated oscillation maps under Earth matter effects demonstrate that the proposed method achieves accurate reconstruction of the six oscillation parameters with empirical coverage closely matching the nominal confidence levels, highlighting its potential as a scalable and interpretable alternative to conventional MC inference.

This contribution represents a first step toward applying map-based inference to real experimental data. 
Here we train and validate our proposed models on simulated oscillation maps that include matter effects but do not model realistic neutrino fluxes nor detector response. 
Extending the pipeline to observed data will require a reconstruction or forward-folding stage that converts detector measurements into oscillation probability maps, a direction we reserve for future work. 
Once that bridge is in place, the models proposed in this work can be applied to reconstructed maps to produce rapid, direct estimates of the underlying oscillation parameters, offering a computationally efficient alternative to MC fits.\footnote{Code and appendix: \url{https://github.com/GiorgioMorales/NuOscParam}}

\section{Related Work} \label{sec:related}

Parameter inference from neutrino oscillation data has long relied on likelihood‐based or global‐fit methods. 
In particular, Markov Chain MC (MCMC) techniques have been widely used to explore the posterior or confidence regions by comparing observed event spectra to simulated expectations~\cite{Hannestad2007}. 
These global‐fit approaches are robust and interpretable, but require expensive repeated forward simulations and likelihood evaluations. 
For instance, Esteban \textit{et al.}~\cite{Esteban2017} performed a combined global fit over reactor, accelerator, solar, and atmospheric data, employing MC sampling to derive allowed regions for oscillation parameters. 
The NOvA long-baseline experiment~\cite{Acero2024} was also reanalyzed using a Bayesian MCMC framework, expanding the inference to include reactor mixing angle constraints and computing full posterior distributions.

Recently, machine learning techniques have been successfully applied to oscillation parameter estimation in an experimental setting~\cite{Abbasi2025}. 
In that study, 9.3 years of atmospheric neutrino data from the IceCube DeepCore detector were analyzed using convolutional neural networks to reconstruct five event-level quantities, including neutrino energy and incoming zenith angle, from photomultiplier readouts.
These results were then binned into 2-D histograms of reconstructed energy versus reconstructed zenith, which served as the observables for the physics fit. 
The actual oscillation parameter estimation, however, was performed entirely through MC.
As such, simulated events were reweighted according to oscillation probabilities for different hypotheses, and the resulting MC templates were statistically compared with the data histograms to extract the best-fit values of the mass splitting $\Delta m^2_{32}$ and mixing angle $\sin^2\theta_{23}$.
A similar analysis was reported by KM3NeT/ORCA using reconstructed energy–zenith templates and a binned likelihood fit on the ORCA6 dataset~\cite{KM3NeT_collaboration2024-jz}. 

Likelihood‐free and simulation‐based inference techniques have also been explored in the neutrino domain.
For instance, Pina‐Otey \textit{et al.}~\cite{PinaOtey2020} proposed a likelihood-free method based on neural spline flows to infer neutrino oscillation parameters in the T2K experiment.
Their input observables are the reconstructed momentum and direction of the outgoing $\mu$ lepton from charged-current interactions.
Using simulated pairs of these observables and oscillation parameters, a normalizing flow model was trained to approximate the joint probability distribution between data and parameters.
Once trained, the model enabled direct estimation of the posterior distribution over oscillation parameters given observed events, eliminating the need for explicit likelihood formulations.

To the best of our knowledge, no prior work has attempted to infer $\nu$ oscillation parameters \textbf{directly from $\nu$ oscillation probability maps}; i.e., full 2-D maps across energy and zenith angle. 
Existing machine learning efforts in neutrino physics typically focus on auxiliary tasks, such as event reconstruction, classification, background suppression, or parameter estimation from reduced observables~\cite{Peterson:2023ayg,Heyer:2024nkq}. 
These works do not learn mappings from structured oscillation probability maps to the fundamental oscillation parameters themselves.

\section{Problem Statement}

Atmospheric neutrino flavor oscillations arise from the mismatch between the flavor and $\nu$ mass eigenstates. 
Our goal is to estimate the parameters governing such oscillations from simulated oscillation probability maps.
Each simulation is parameterized by a vector $\mathbf{p} = \left[ \theta_{12}, \theta_{23}, \theta_{13}, \delta_{\mathrm{CP}}, \Delta m_{21}^2, \Delta m_{31}^2 \right]$,
where $\theta_{12}, \theta_{23}, \theta_{13}$ are the leptonic mixing angles (in degrees), $\delta_{\mathrm{CP}}$ is the CP-violating phase (in degrees), and $\Delta m_{21}^2, \Delta m_{31}^2$ are the squared mass differences (in $\mathrm{eV}^2$). 

Given $\mathbf{p}$, a simulator $\mathcal{S}$ produces an oscillation probability map $\mathcal{S}(\mathbf{p})=\mathbf{X} \in \mathbb{R}^{9 \times H \times W}$, which encodes the transition probabilities among neutrino flavors. 
Neutrino flavors, denoted by $\nu_e, \nu_\mu, \nu_\tau$, correspond to the three leptonic states associated respectively with the electron, muon, and tau leptons~\cite{PDG2024}. 
For each ordered pair $(\alpha,\beta) \in \{e,\mu,\tau\}^2$, the channel $\nu_\alpha \to \nu_\beta$ represents the probability that a neutrino produced in flavor state $\nu_\alpha$ is later detected as flavor $\nu_\beta$. 
The nine channels, indexed by $c \in \{1,\dots,9\}$, are ordered such that the $c$-th channel corresponds to the transition $\nu_{\alpha_i} \to \nu_{\alpha_j}$, where $c = 3(i-1) + j$ with $i,j \in \{1,2,3\}$ and $(\alpha_1,\alpha_2,\alpha_3) = (e,\mu,\tau)$.

The vertical axis of each channel, of size $H$, corresponds to discretized values of $\cos\theta$, where $\theta$ is the zenith angle of the neutrino trajectory with respect to the detector. 
It determines the path length traveled by the neutrino through the Earth, which strongly influences oscillation behavior~\cite{Fogli1998,Kajita}. 
The horizontal axis, of size $W$, represents discretized neutrino energies $E$. 
Thus, each entry in a given channel of the map $\mathbf{X}$ represents the oscillation probability for a given transition $\nu_\alpha \to \nu_\beta$ at a specific pair $(E, \cos\theta)$.

Let $\mathbf{X}^{\text{in}}\in\mathbb{R}^{C\times H\times W}$ denote the input map used for inference with $C\le 9$; that is, a subset of the nine flavor-transition channels may be selected depending on task-specific considerations.
The general objective is to recover the parameter vector $\mathbf{p}$ from $\mathbf{X}^{\text{in}}$. 
This can be posed as a joint, multi-parameter estimation problem that seeks to learn a mapping $f:\mathbb{R}^{C\times H\times W}\mapsto\mathbb{R}^6$ with learnable parameters $\Theta$, such that $f_{\Theta}(\mathbf{X}^{\text{in}}, \Theta)\approx\mathbf{p}$.
Alternatively, the problem can be posed as a collection of task-specific scalar estimation problems, which, for each parameter $p^{(i)} \in \mathbf{p}$, seek a mapping $f^{(i)}:\mathbb{R}^{C\times H\times W}\mapsto\mathbb{R}$ with parameters $\Theta_i$, such that $f^{(i)}(\mathbf{X}^{\text{in}}, \Theta_i) \approx p^{(i)}$.
In this work, we adopt the latter formulation, treating each oscillation parameter as an independent regression target. 
This decision is formally justified in Appendix~\ref{app:A}.

\section{Materials and Methods}

\subsection{Oscillation Maps Simulators}
\label{sec:data}

For a given parameter set $\mathbf{p}$, we generate oscillation-probability maps that express atmospheric $\nu$ transition probabilities as functions of $E$ and $\cos\theta$. The simulator $\mathbf{X}=\mathcal{S}(\mathbf{p})$ follows the standard three-flavor formalism~\cite{PDG2024}, summarized below.
We consider the restricted parameter ranges reported in Table~\ref{tab:param_ranges}, which are consistent with global-fit constraints~\cite{Esteban2020}.

\begin{table}[t]
\centering
\caption{Operating ranges of the $\nu$ oscillation parameters}
\label{tab:param_ranges}
\renewcommand{\arraystretch}{1.}
\setlength{\tabcolsep}{5pt}
\scriptsize
\begin{tabular}{lcc}
    \toprule
    \textbf{Parameter} & \textbf{Range} & \textbf{Midpoint} \\
    \midrule
    $p^{(1)} = \theta_{12}$ & $[31.27^\circ,\; 35.86^\circ]$ & $33.565^\circ$ \\
    $p^{(2)} = \theta_{23}$ & $[40.1^\circ,\; 51.7^\circ]$ & $45.9^\circ$ \\
    $p^{(3)} = \theta_{13}$ & $[8.20^\circ,\; 8.94^\circ]$ & $8.57^\circ$ \\
    $p^{(4)} = \delta_{\mathrm{CP}}$ & $[120^\circ,\; 369^\circ]$ & $244.5^\circ$ \\
    $p^{(5)} = \Delta m_{21}^2$ & $[6.82,\; 8.04] \times 10^{-5}\,\mathrm{eV}^2$ & $7.43 \times 10^{-5}\,\mathrm{eV}^2$ \\
    $p^{(6)} = \Delta m_{31}^2$ & $[2.431,\; 2.599] \times 10^{-3}\,\mathrm{eV}^2$ & $2.515 \times 10^{-3}\,\mathrm{eV}^2$ \\
    \bottomrule
\end{tabular}
\vspace{-2ex}
\end{table}

\subsubsection{Vacuum Oscillation Maps}

Given $\mathbf{p}$, the standard PMNS matrix $U =
R_{23}(\theta_{23})\,
\Gamma_{\delta_{\mathrm{CP}}}\,
R_{13}(\theta_{13})\,
R_{12}(\theta_{12})
$
is constructed, where $R_{ij}(\theta_{ij})$ denotes a real rotation in the $(i,j)$ plane that mixes mass states $i$ and $j$, and $\Gamma_{\delta_{\mathrm{CP}}}
=\mathrm{diag}(1,\,1,\,e^{i\delta_{\mathrm{CP}}}).$
For a neutrino of energy $E$ incident with zenith $\theta$, the propagation distance through vacuum is given by
$L(\cos\theta) = 2\,R_{\oplus}\cos\theta$,
with $R_{\oplus}=6386~\mathrm{km}$.
The corresponding mass-eigenstate phase factors are
$
\Phi_{i}(E,\cos\theta)
=
\exp\!\left(
-\frac{i}{2E}\,
\Delta m_{i1}^{2}\,
L(\cos\theta)
\right).$
The transition probabilities are obtained from the full three-flavor
quantum evolution operator,
$P^{(\mathrm{vacuum})}_{\alpha\rightarrow\beta}(E,\cos\theta)
= | [ U\, \Phi(E,\cos\theta)\, U^{\top}]_{\beta\alpha} |^{2}$,
evaluated on a $(E,\cos\theta)$ pair.

\subsubsection{Matter-Effect Oscillation Maps}

When neutrinos traverse the Earth, coherent forward scattering on electrons modifies their flavor evolution. 
We simulate this by combining the standard three-flavor mixing formalism with a radial Earth density profile given by the Preliminary Reference Earth Model (PREM)~\cite{Dziewonski1981}. 
For each $(E,\cos\theta)$ pair, we compute the path through the Earth layers and propagate the neutrino flavor state using a piecewise constant-density approximation. 
The matter potential, implemented following the Mikheev–Smirnov–Wolfenstein (MSW) formalism~\cite{PDG2024,Wolfenstein1978}, is incorporated directly into the Hamiltonian governing neutrino evolution. 
This produces flavor-transition probabilities $P^{(\mathrm{matter})}_{\alpha\rightarrow\beta}(E,\cos\theta)$ that fully account for matter effects.

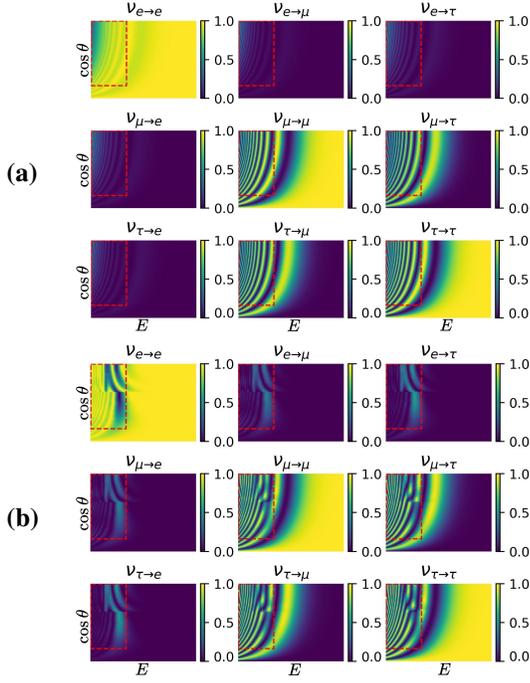
\begin{figure}[t] 
    \centering
    \begin{subfigure}{1.0\columnwidth}
        \centering
        \setlength{\unitlength}{1cm}
        \begin{picture}(0,0)
            \put(-0.8,2){\textbf{(a)}}
        \end{picture}
        \includegraphics[width=0.68\textwidth]{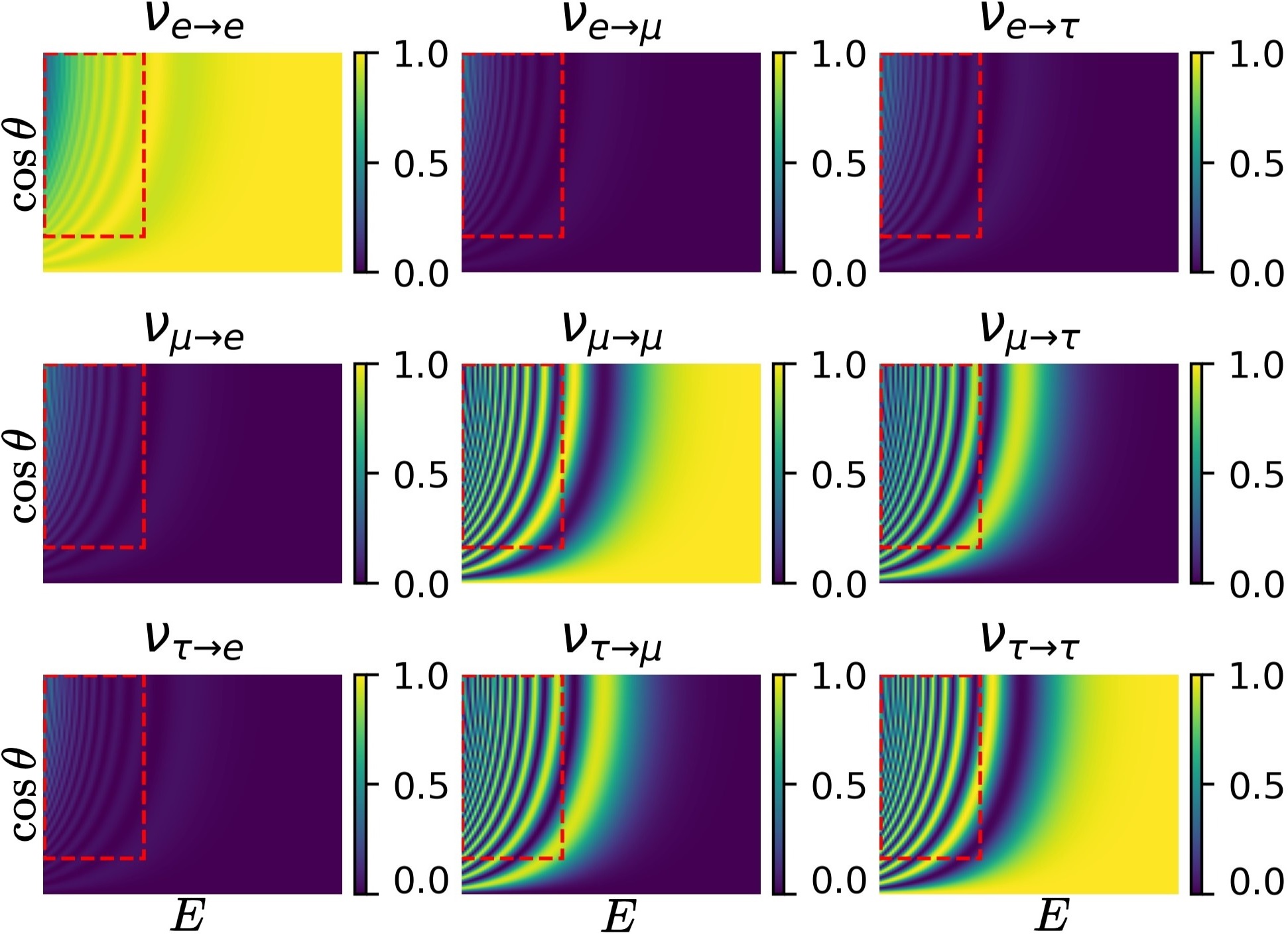}
    \end{subfigure}
    \vspace{-1.5ex}

    \begin{subfigure}{1.0\columnwidth}
        \centering
        \setlength{\unitlength}{1cm}
        \begin{picture}(0,0)
            \put(-0.8,2){\textbf{(b)}}
        \end{picture}
        \includegraphics[width=0.68\textwidth]{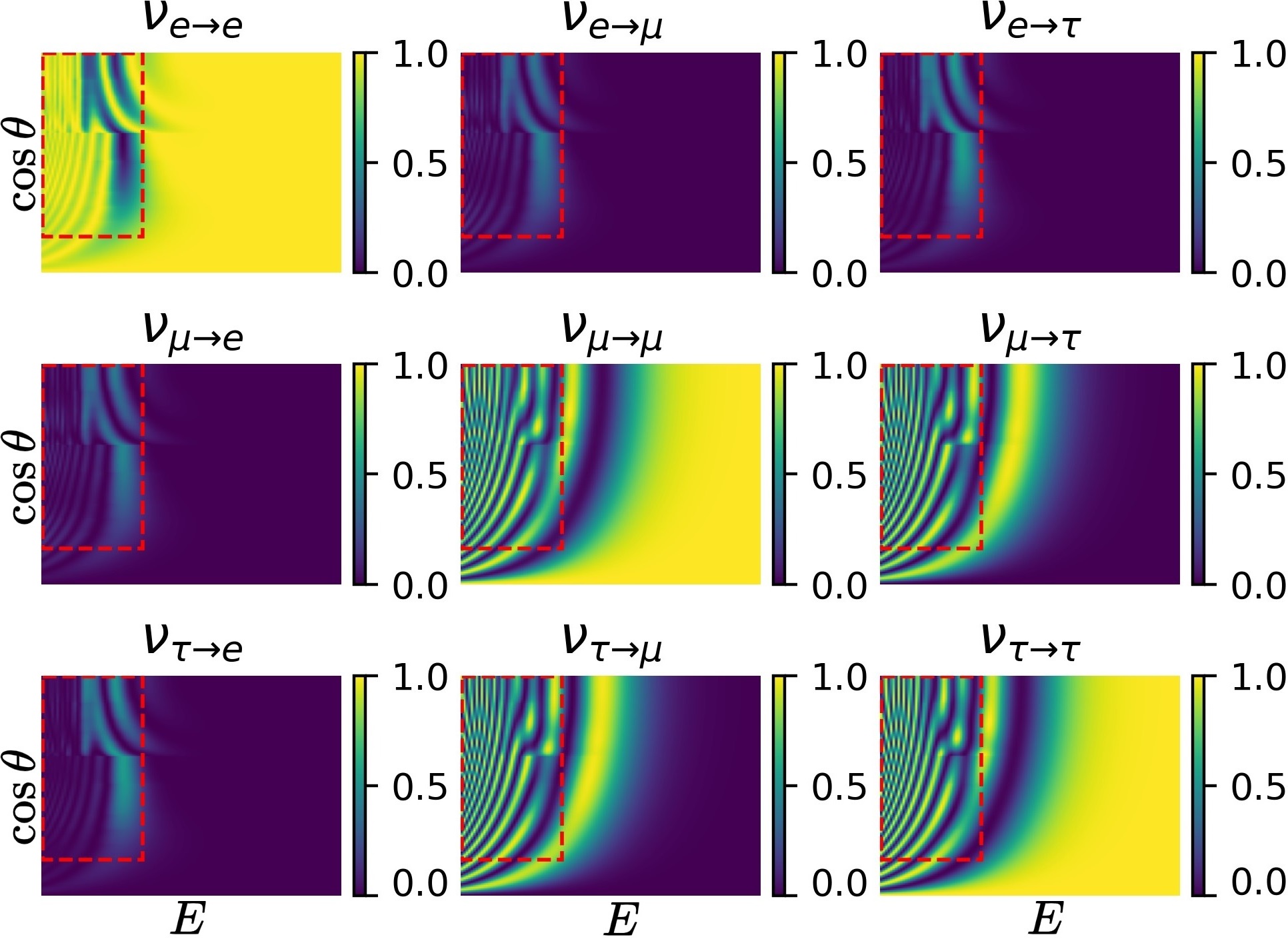}
    \end{subfigure}
    \vspace{-3.5ex}
    \caption{$\nu$ oscillation maps in \textbf{(a)} vacuum, \textbf{(b)} after matter effect for $\mathbf{p} = [359.9, 410.5, 8.37, 156.7, 6.59 \!\times\! 10^{-5}, 2.65 \!\times\! 10^{-3}]$.}
    \label{fig:maps}
    \vspace{-2ex}
\end{figure}

Both vacuum and matter-effect oscillation maps are generated by evaluating the transition probabilities $P^{(\mathrm{vacuum})}_{\alpha\rightarrow\beta}(E,\cos\theta)$ and $P^{(\mathrm{matter})}_{\alpha\rightarrow\beta}(E,\cos\theta)$ on a discrete grid $\{(E_i, \cos\theta_j)\}_{i,j=1}^{120}$.
Since this work focuses on atmospheric neutrinos, we consider energies $E_i$ logarithmically spaced in $[1, 10^3]$~GeV and $\cos\theta_j$ linearly spaced in $[0, 1]$, yielding $120 \times 120$ probability maps.
Fig.~\ref{fig:maps} depicts a comparison of the maps generated in vacuum and after matter effects for the same $\mathbf{p}$. 
To reduce computational cost, we focus only on the first 80 rows and 30 columns (highlighted in red), which contain the majority of the oscillation information.


\subsubsection{Channel Selection}

The nine-channel maps exhibit parameter-dependent sensitivity.
To enhance identifiability and minimize redundancy, we select the three most discriminative channels ($C=3$) for each parameter, based on known oscillation behavior and their response to controlled perturbations.

\begin{itemize}[leftmargin=1em, noitemsep,topsep=0pt]
    \item $\theta_{12}$: Varying $\theta_{12}$ modifies the mixing between the first two mass eigenstates, producing the broadest long-wavelength oscillation pattern in all flavors, which is clearly visible in the survival channels $(e\!\to\!e)$, $(\mu\!\to\!\mu)$, and $(\tau\!\to\!\tau)$.
    \item $\theta_{23}$: The channels $(\mu\!\to\!\mu)$, $(e\!\to\!\mu)$, and $(e\!\to\!\tau)$ are chosen because $\theta_{23}$ controls $\nu_\mu$–$\nu_\tau$ mixing.
    Its effect appears most clearly in $\nu_\mu\!\to\!\nu_\mu$ as atmospheric-frequency oscillation depth changes. The appearance channels $\nu_e\!\to\!\nu_\mu$ and $\nu_e\!\to\!\nu_\tau$ encode how electron-flavor amplitude is split between $\mu$ and $\tau$, a pattern strongly governed by $\theta_{23}$.
    \item $\theta_{13}$: The channels $(e\!\to\! e)$, $(e\!\to\!\mu)$, and $(e\!\to\!\tau)$ are the most sensitive to $\theta_{13}$, which directly shapes $\nu_e\!\to\!\nu_e$ disappearance at atmospheric frequencies and controls the amplitudes of $\nu_e\!\to\!\nu_\mu$ and $\nu_e\!\to\!\nu_\tau$ appearance. Because its imprint is concentrated in the electron sector, the three electron-origin channels provide maximal sensitivity.
    \item $\delta_{\rm CP}$: Channels $(e\!\to\! \mu)$, $(e\!\to\!\tau)$, and $(\tau\!\to\!e)$ are selected. The CP phase enters through interference terms that predominantly affect appearance probabilities. The channels $\nu_e\!\to\!\nu_\mu$ and $\nu_e\!\to\!\nu_\tau$ contain the strongest CP-dependent distortions, while $\nu_\tau\!\to\!\nu_e$ adds complementary interference structure that helps separate CP effects. 
    \item $\Delta m_{21}^2$: Channels $(e\!\to\! e)$, $(e\!\to\!\mu)$, and $(e\!\to\!\tau)$ are selected. $\Delta m_{21}^2$ produces slow oscillation modulations that appear most prominently in electron-sector probabilities at low energies. These broad features manifest clearly in $\nu_e!\to!\nu_e$ and propagate into the electron-appearance channels.
    \item $\Delta m_{31}^2$: Channels $(\mu\!\to\!\mu)$, $(e\!\to\!e)$, and $(\tau\!\to\!\tau)$ are selected. $\Delta m_{31}^2$ sets the dominant atmospheric-frequency oscillation scale and generates sharp energy-dependent structures in all survival channels. $\nu_\mu\!\to\!\nu_\mu$ carries the strongest signal, while the other channels encode complementary frequency information modulated by their respective mixing angles.
\end{itemize}

\subsection{Structured Hierarchical Transformer}

We introduce a hierarchical architecture that respects the structure of the oscillation maps explicitly. 
Recall that each column of the input $\mathbf{X}^{\text{in}}$ corresponds to the angular profile at a fixed energy, while variation across columns encodes the energy dependence of these profiles. This configuration naturally suggests a two-level hierarchy. 
To exploit it, we propose: First, an energy-specific encoder that processes the angular structure of each column, and, second, a global encoder that models how these angular features vary across energies.

Consider the model $f^{(i)}$, trained to estimate parameter $p^{(i)}$.
Formally, let $\mathbf{x}^{(k)} = \bigl(\mathbf{X}^{\text{in}}_{:,1,k}, \mathbf{X}^{\text{in}}_{:,2,k}, \dots, \mathbf{X}^{\text{in}}_{:,H,k}\bigr)\in\mathbb{R}^{H\times C}$ denote the column at the $k$-th energy bin, which can be interpreted as a sequence representing the transition probabilities at different $\cos\theta$ values. 
Each sequence is passed to an energy-specific inner transformer $\mathcal{T}_k$, yielding $\mathbf{u}^{(k)} = \mathcal{T}_k\bigl(\mathbf{x}^{(k)}\bigr) \in \mathbb{R}^{H\times d_{\mathrm{ang}}}$, where $d_{\mathrm{ang}}$ is the embedding dimension. 
A pooling operator then compresses the angular dimension, producing a per-energy embedding $\mathbf{e}_k = \operatorname{Pool}\bigl(\mathbf{u}^{(k)}\bigr) \in \mathbb{R}^{d_{\mathrm{ang}}}$.
Stacking all $W$ embeddings yields the sequence of encoded angular features $\mathbf{S} = [\mathbf{e}_1, \dots, \mathbf{e}_W] \in \mathbb{R}^{W\times d_{\mathrm{ang}}}$.

The energy-level sequence $\mathbf{S}$ is then processed by an outer transformer encoder $\mathcal{T}_{\mathrm{outer}}$, producing $\mathbf{H} = \mathcal{T}_{\mathrm{outer}}(\mathbf{S}) \in \mathbb{R}^{W\times d_{\mathrm{en}}}$, followed by a global pooling to obtain a single representation $\mathbf{h} = \operatorname{Pool}\bigl(\mathbf{H}\bigr) \in \mathbb{R}^{d_{\mathrm{en}}}$.
Finally, a fully-connected layer $L$ produces the scalar regression output $p^{(i)}$:
$\hat{p}^{(i)} = f^{(i)}\bigl(\mathbf{X}^{\text{in}}\bigr) = L(\mathbf{h}).$
In addition, we include low-amplitude sinusoidal positional encodings along both the energy and angular axes after feature projection, thereby providing the transformers with explicit coordinate information.

Our two-stage design is physically motivated: inner encoders operate at fixed energy and capture correlations across the discrete $\cos\theta$ bins that define the angular dependence of oscillation probabilities. 
The outer encoder then integrates these per-energy embeddings to learn how angular signatures vary across energy, thus modeling global spectral relationships. 
Self-attention is particularly suitable, since oscillation patterns may involve extended correlations in $\cos\theta$ and nonlocal dependencies across energies. 
Thus, attention aggregates information across the full angular domain and across widely separated energies without enforcing strict locality.

Furthermore, unlike other hierarchical transformers~\cite{chalkidis-etal-2022-hat} that reuse a single inner encoder with shared weights across segments, the inner encoders $\{\mathcal{T}_j\}_{j=1}^W$ are independent in this approach. 
Weight sharing assumes that segments are functionally equivalent, whereas, in this problem, each column corresponds to a distinct energy bin with different oscillatory behavior. 
Treating them as stationary is unwarranted.
Our architecture also bears similarity to separable transformers~\cite{sepTR}, which use two sequential transformer blocks that attend along different axes of the input. 
However, they rely on a single transformer per axis to learn relationships that are assumed to hold uniformly across all positions of that axis.
By contrast, decoupling angular encoding across energies allows modeling energy-specific angular structures while still learning coherent cross-energy interactions at the outer level.

More general alternatives, such as 2-D attention or flattened-sequence transformers, are either computationally prohibitive or obscure fine-grained oscillation features by prematurely entangling angular and spectral structure. 
Similarly, Vision Transformers risk discarding important subtle variations in the oscillation maps, as further discussed in Section~\ref{discussion}.

\subsection{Simulation-augmented Physics-Aware Neural Network}
\label{sec:training}

A key difficulty in oscillation parameter estimation is that close parameter values can correspond to nearly indistinguishable oscillation maps. 
Optimizing solely on the predicted parameter error is therefore not sufficient, as small parameter deviations may lead to gradients that fail to distinguish informative directions for improvement. 
To address this, we use a surrogate consistency step~\cite{sim_reg,besnard:hal-05066556}, in which the predicted parameter must not only approximate the target but also regenerate an oscillation map consistent with the input. 
This reconstruction requirement accentuates differences between the estimated and ground-truth parameters, since even subtle changes in parameter values can propagate into structural variations in the simulated maps. 
By enforcing consistency in both parameter and simulation space, the model receives a richer training signal that improves optimization and yields estimates that are both accurate and physically meaningful.

Formally, let us consider the input $\mathbf{X}^{\text{in}}$, generated from the ground-truth vector $\mathbf{p}$, and the model $f^{(i)}$ that produces $\hat{p}^{(i)}$.
This prediction is injected into the full oscillation parameter vector, yielding the vector $\mathbf{\hat{p}}$ (i.e., $\mathbf{\hat{p}}[i] = \hat{p}^{(i)}$ and $\mathbf{\hat{p}}[t] = \mathbf{p}[t]$ for $t \neq i$).
Then, a reconstructed map is produced using a set of pretrained, differentiable surrogate models, $\mathbf{X}^{\text{sim}} = \mathcal{\hat{S}}(\mathbf{\hat{p}})$, where $\mathcal{\hat{S}}$ denotes a composition of per-channel models.
In this work, we employ feedforward neural networks as the surrogate models. 
Further architectural and training details about these surrogate models are provided in Appendix~\ref{app:B}.

As such, model $f^{(i)}$ is optimized with a two-term objective. 
The primary term is the mean squared error (MSE) on the parameter space, $\mathcal{L}^{(i)}_{\mathrm{par}} = \frac{1}{N} \sum_{j=1}^{N} \big(\hat{p}^{(i)}_j - p^{(i)}_j\big)^2$, where $N$ is the batch size. 
The auxiliary term is a reconstruction loss enforcing map consistency, $\mathcal{L}^{(i)}_{\mathrm{rec}} = \sum_{j=1}^{N} \big\| \mathbf{X}^{\text{in}}_j - \mathbf{X}^{\text{sim}}_j \big\|_F^2$.
The total loss is $\mathcal{L}^{(i)}_{\mathrm{total}} = \mathcal{L}^{(i)}_{\mathrm{par}} + \lambda \mathcal{L}^{(i)}_{\mathrm{rec}}$, with $\lambda$ a scheduled weight balancing the two terms.
Thus, $\Theta_i = \arg\min_{\Theta_i} \ \mathcal{L}^{(i)}_{\mathrm{total}}$.
This corresponds to a Simulation-augmented Physics-Aware Neural Network (SimPANN) approach~\cite{besnard:hal-05066556}, in which physics is incorporated through differentiable surrogate simulations rather than explicit analytical equation residuals, distinguishing it from Physics-Informed Neural Networks (PINNs).

This approach also shares similarity with Simulation-Based Inference (SBI)~\cite{SIB}, since $\mathcal{\hat{S}}(\mathbf{p})$ maps oscillation parameters to oscillation maps without an analytically tractable likelihood. 
Our models learn amortized inference functions that recover latent parameters from simulated maps--parameter pairs, bypassing explicit likelihood evaluation.
This is further reinforced by the surrogate simulator consistency step, which enforces that the inferred parameters, when re-simulated through differentiable surrogates, reproduce the input oscillation maps, closing the loop between inference and simulation.
This interpretation is supported by the uncertainty quantification method addressed in the following section, which estimates the confidence of inferred parameters in a likelihood-free setting.

\subsection{Uncertainty Quantification} \label{sec:UQ}
\vspace{-0.3ex}

Reliable uncertainty estimation is crucial in oscillation parameter inference, as it provides a rigorous assessment of prediction confidence. 
To this end, we leverage DualAQD~\cite{10365540}, a training framework with a specialized loss function that, alongside the point-prediction models $f^{(i)}$, introduces auxiliary networks dedicated to producing prediction intervals (PIs).

The network that predicts PIs for the $i$-th oscillation parameter, denoted by $f^{(i)}_{\text{PI}}$, takes as input the full set of estimated oscillation parameters $\mathbf{\hat{p}} = \{ \hat{p}^{(1)},\dots, \hat{p}^{(6)}\}$ for a given map $\mathbf{X}$ and outputs the lower and upper bounds $[\hat{\ell}^{(i)}, \, \hat{u}^{(i)}] = f^{(i)}_{\text{PI}}(\mathbf{\hat{p}};\boldsymbol{\Theta}_{f^{(i)}_{\text{PI}}})$,
where $\boldsymbol{\Theta}_{f^{(i)}_{\text{PI}}}$ denotes its trainable weights.
Conditioning the PI models on $\mathbf{\hat{p}}$ is justified since the operating domain of the oscillation parameters is deliberately restricted and non-periodic (Sec.~\ref{sec:data}), making each parameter combination effectively unique.
Thus, the vector $\mathbf{\hat{p}}$ acts as a highly informative, low-dimensional summary of the physical state that produced the observation.
Using $\mathbf{\hat{p}}$ as input is also computationally attractive as it reuses already computed point estimates and decouples PI learning from the expensive forward models relative to raw oscillation maps.

The parameters of $f^{(i)}_{\text{PI}}$ are then optimized by minimizing $
\boldsymbol{\Theta}_{f^{(i)}_{\text{PI}}} = {\arg\min}_{\boldsymbol{\Theta}_{f^{(i)}_{\text{PI}}}} \ \mathcal{L}_1 + \beta \, \mathcal{L}_2,
$
known as the DualAQD loss, where $\mathcal{L}_1 = \frac{1}{N} \sum_{j=1}^{N} \big(|\hat{u}^{(i)}_j - p^{(i)}_j| + |p^{(i)}_j - \hat{\ell}^{(i)}_i|\big)$ penalizes excessively wide intervals for captured samples, and $\mathcal{L}_2 = e^{\xi - d_u} + e^{\xi - d_\ell}$ encourages interval integrity, ensuring that $\hat{\ell}^{(i)} \leq \hat{p}^{(i)} \leq \hat{u}^{(i)}$ on average. 
Here, $d_u = \tfrac{1}{N}\sum_{j=1}^N (\hat{u}^{(i)}_j - \hat{p}^{(i)}_j)$ and $d_\ell = \tfrac{1}{N}\sum_{i=1}^N (\hat{p}^{(i)}_j - \hat{\ell}^{(i)}_j)$, while $\beta$ is a self-adaptive coefficient balancing $\mathcal{L}_1$ and $\mathcal{L}_2$.  
This allows producing narrow PIs with high empirical coverage without compromising accuracy.

While DualAQD's optimization-based approach produces PIs with good empirical coverage and narrow width, it lacks a formal mechanism to guarantee coverage across arbitrary data distributions. 
We address this limitation by introducing \emph{Conformal DualAQD}, which augments the initial PIs generated using NNs with a split-conformal wrapping step. 

Our setting employs a distinct $N_c$-sample calibration set,
independent of the training data, to determine the adjustment needed to achieve the target coverage.
For each calibration point $(\mathbf{X}^{\text{in}}_j, \hat{p}^{(i)}_j)$, we compute a nonconformity score based on the one-sided violation of the DualAQD interval: $s_j^{(i)} \!=\! {(\max\{p^{(i)}_j - \hat{u}^{(i)}_j,\; \hat{\ell}^{(i)}_j - p^{(i)}_j,\; 0\})}/({w_j^{(i)} + \varepsilon})$, $j\!=\!1,\dots,n$,
where $w_j^{(i)} \!=\! \hat{u}^{(i)}_j - \hat{\ell}^{(i)}_j$ and $\varepsilon$ is a vanishing regularizer introduced to ensure numerical stability.
This width normalization enforces scale invariance, preventing undue penalization of naturally wider intervals, as in heteroskedastic settings~\cite{nonconformity}.

We then set $q_\alpha^{(i)}$ to be the $\lceil (N_c+1)(1-\alpha)\rceil$-th smallest score in the calibration set. 
During inference, the conformally calibrated interval for a new input is produced by augmenting the original DualAQD interval by $q_\alpha^{(i)} w$, which produces a multiplicative conformal correction that preserves the asymmetry of the base interval: $[\hat{\ell}^{(i)} - q_\alpha^{(i)} w,\; \hat{u}^{(i)} + q_\alpha^{(i)} w].$
Under the standard exchangeability assumption~\cite{angelopoulos2025}, this procedure guarantees marginal coverage at the $1-\alpha$ level.

\section{Experimental Results} \label{sec:results}

We evaluated our approach on a dataset of $\nu$ oscillation maps with Earth matter effects. 
In particular, we synthesized three independent data collections using the simulator $\mathcal{S}$:
\begin{itemize}[leftmargin=1em, noitemsep,topsep=0pt]
\item A \emph{training/validation} dataset of 20,000 samples. We use 80\% for training and the remaining 20\% as a validation split to select the best-performing weights for each model.
\item A \emph{calibration} dataset of 10,000 independent samples used for the split-conformal calibration of the DualAQD PIs.
\item Ten mutually independent \emph{test} sets, each of size 1,000, used for baseline comparison and reporting of uncertainties. This was chosen for practicality, as the baseline method was computationally unfeasible to run on sets of 10,000 samples.
\end{itemize}

For any data collection, the $j$-th sample is generated by first sampling a ground-truth oscillation parameter vector $\mathbf{p}_j$ uniformly within the ranges listed in Table~\ref{tab:param_ranges}.
From it, the selected simulator produces the oscillation maps $\mathbf{X}_j = \mathcal{S}(\mathbf{p}_j)$.

For the model architecture of the $f^{(i)}$ models, due to the high computational expense associated with training a single model, we used a one-factor-at-a-time approach to choose the hyperparameters: the number of layers of $\mathcal{T}_k$, 8, the number of layers of $\mathcal{T}_\mathrm{outer}$, 8, the number of heads of all transformer layers, 16, and the embedding dimensions $d_{\mathrm{ang}}\!=\!256$ and $d_{\mathrm{en}}\!=\!256$.
In addition, the balancing weight $\lambda$ used in the transformer's objective function follows an exponential decay schedule that progressively down-weights $\mathcal{L}_{\mathrm{rec}}$ during training, increasing emphasis on accurate parameter estimation in the later training stages.
On the other hand, the PI-estimation models $f_{\mathrm{PI}}^{(i)}$ are designed as feed-forward NNs with two hidden layers, each with 32 nodes with ReLU activation.

Once calibration is complete (Sec.~\ref{sec:UQ}), all trained models and calibration factors are used to perform inference on each of the ten test sets independently. 
For each parameter $p^{(i)}$ and each test set of size $N_T=1000$, we compute the root-mean-square error $\mathrm{RMSE}^{(i)} = \sqrt{\frac{1}{N_T} \sum_{j=1}^{N_T}(\hat{p}_j^{(i)} - p_j^{(i)})^2}$.
We also report the relative RMSE, defined as $\mathrm{RMSE}^{(i)} / m^{(i)}$, where $m^{(i)}$ is the midpoint of the operating range for parameter $p^{(i)}$.

The PI estimation models $f^{(i)}_{\text{PI}}$ and the subsequent conformal calibration step were performed using a nominal miscoverage rate of $\alpha = 0.1$, corresponding to 90\% PIs, a common choice in physics applications.
To assess the performance of our uncertainty quantification approach, (the PI estimation models and the calibration step were carried out considering $\alpha=0.1$, as it is commonly used in physics applications, verify if this is accurate), we calculate the mean prediction-interval width $\mathrm{MPIW}^{(i)} = \frac{1}{N_T}\sum_{j=1}^{N_T} \bigl(\hat{u}^{(i)}_j - \hat{\ell}^{(i)}_j\bigr)$, and the prediction interval coverage probability $\mathrm{PICP}^{(i)} = \frac{1}{N_T}\sum_{j=1}^{N_T} \mathbf{1}\bigl\{\hat{\ell}^{(i)}_j \le p^{(i)}_j \le \hat{u}^{(i)}_j\bigr\}$ that represents the empirical fraction of true values captured by the conformalized PIs.
We aim to achieve high coverage (i.e., $\mathrm{PICP}^{(i)} \approx 1 - \alpha$) with minimal PI width.

In addition, we implemented a delayed-acceptance MCMC procedure~\cite{accMCMC,accMCMC2} as a baseline method for comparison.  
Given an observed oscillation map $\mathbf{X}$, MCMC first constructs a chain of parameter proposals $\{\mathbf{p}^{(t)}\}$ using the surrogate simulator $\hat{\mathcal{S}}$.  
In particular, at iteration $t$, a candidate $\mathbf{p}^\star$ is drawn from a symmetric proposal kernel; its corresponding surrogate-generated map $\hat{\mathbf{X}}^\star = \hat{\mathcal{S}}(\mathbf{p}^\star)$ is compared to the observed map to compute approximate likelihoods.
At the final stage, the surrogate chain is corrected using a delayed-acceptance step. 
The algorithm evaluates the exact simulator $\mathcal{S}$ on proposed samples,  within a fixed budget, and applies a Metropolis–Hastings acceptance step using the exact log-likelihoods.
The MCMC hyperparameters were chosen via preliminary tuning on a validation set to balance efficient exploration of the parameter space and the computational cost of full simulations.
The number of walkers was set to 36, the number of steps to 800, burn-in to 200, initial spread to 0.02, and a budget of 200 evaluations for the exact simulator, with a probability of using the exact simulator at each step of 0.05. 

The MCMC posterior samples provide 90\% credible intervals (CIs), $[\ell^{(i)}_{\mathrm{CI}}, u^{(i)}_{\mathrm{CI}}]$, for each $\nu$ parameter.
To transform CIs into PIs, $[\ell^{(i)}, u^{(i)}]$, we apply a conformal calibration step that adds a residual-based nonconformity margin.
In particular, we estimate the empirical residual dispersion on the calibration set, $\sigma^{(i)}=\mathrm{std}(\mathbf{p}^{(i)} - \hat{\mathbf{p}}^{(i)})$, and inflate the bounds by a parameter-specific factor $c^{(i)}$ such that $[\ell^{(i)}, u^{(i)}] = [\ell^{(i)}_{\mathrm{CI}} - c^{(i)} \sigma^{(i)}, u^{(i)}_{\mathrm{CI}} + c^{(i)} \sigma^{(i)}]$.
The factors $c^{(i)}$ are chosen such that the resulting intervals achieve 90\% empirical coverage.
 
The performance metrics (i.e., $\mathrm{RMSE}^{(i)}$, Rel. $\mathrm{RMSE}^{(i)}$, $\mathrm{MPIW}^{(i)}$, and $\mathrm{PICP}^{(i)}$) are computed per test set. 
Table~\ref{tab:results_earth} reports, for each parameter, the mean metric values and their corresponding standard deviation across the ten sets. 
The bold entries indicate the method that achieved the best performance value and that its difference with respect to the values obtained by the other method is statistically significant according to a paired $t$-test performed at the 0.05 level. 
The distribution of the metrics obtained by both methods is depicted in Appendix~\ref{app:C} using box plots.
Table~\ref{tab:cost} compares the computational cost of the two approaches in terms of MFLOPs and average wall-clock time.  
All measurements were taken under comparable conditions,
with both methods using single-GPU acceleration (A100-80GB GPU, AMD 7763 64-Core CPU, 500 GB RAM).

This comparison should not be interpreted as one conducted under identical training objectives. 
While MCMC does not rely on parameter-space supervision and incurs its full computational cost at inference time, our approach uses an offline, parameter-labeled training phase to learn a direct inverse mapping. 
The comparison is nevertheless fair at inference time: both methods operate solely on oscillation maps and target the same physical parameters. 
Thus, the results highlight the trade-off between per-instance simulation-based inference and amortized inference, where computational cost is shifted from inference to a one-time training stage.

\begin{table*}[t] \centering
\caption{Performance comparison across the six oscillation parameters}
\label{tab:results_earth}
\renewcommand{\arraystretch}{1}
\resizebox{0.88\textwidth}{!}{%
\begin{tabular}{lcccccccc}
\toprule
\multirow{2}{*}{Param.} & \multicolumn{4}{c}{\textbf{Proposed Method}} & \multicolumn{4}{c}{\textbf{MCMC Baseline}} \\[-2pt]
\cmidrule(lr){2-5} \cmidrule(lr){6-9}
& RMSE & \makecell{Rel. RMSE~(\%)} & MPIW & PICP & RMSE & \makecell{Rel. RMSE~(\%)} & MPIW & PICP \\[-2pt]
\midrule
$\theta_{12}$ & $\mathbf{0.0704 \pm 0.0020}$ & $\mathbf{0.210 \pm 0.006}$ & $\mathbf{0.0821 \pm 0.0014}$ & $90.33 \pm 1.00$ & $0.1510 \pm 0.0202$ & $0.450 \pm 0.060$ & $0.2626 \pm 0.0296$ & $88.53 \pm 1.59$ \\
$\theta_{23}$ & $0.0268 \pm 0.0010$ & $0.058 \pm 0.002$ & $\mathbf{0.0225 \pm 0.0011}$ & $89.94 \pm 0.77$ & $0.0273 \pm 0.0024$ & $0.059 \pm 0.005$ & $0.0651 \pm 0.0051$ & $90.37 \pm 0.99$ \\
$\theta_{13}$ & $0.0038 \pm 0.0002$ & $0.044 \pm 0.002$ & $\mathbf{0.0033 \pm 0.0002}$ & $90.12 \pm 0.79$ & $0.0037 \pm 0.0005$ & $0.043 \pm 0.006$ & $0.0089 \pm 0.0009$ & $91.66 \pm 1.46$ \\
$\delta_{\mathrm{CP}}$ & ${0.8877 \pm 0.0272}$ & ${0.363 \pm 0.011}$ & $\mathbf{1.1187 \pm 0.033}$ & $90.17 \pm 0.01$ & $\mathbf{0.7796 \pm 0.1064}$ & $\mathbf{0.319 \pm 0.043}$ & $1.8681 \pm 0.2184$ & $91.26 \pm 1.49$ \\
$\Delta m_{21}^2$ & $\mathbf{(2.89 \pm 0.01)\,\text{E-7}}$ & $\mathbf{0.389 \pm 0.001}$ & $\mathbf{(3.81 \pm 0.00)\,\text{E-7}}$ & $89.89 \pm 1.03$ & $(7.28 \pm 0.40)\,\text{E-7}$ & $0.980 \pm 0.054$ & $(1.67 \pm 0.08)\,\text{E-6}$ & $90.53 \pm 1.12$ \\
$\Delta m_{31}^2$ & $(3.89 \pm 0.01)\,\text{E-7}$ & $0.015 \pm 0.000$ & $\mathbf{(5.43 \pm 0.06)\,\text{E-7}}$ & $89.42 \pm 0.86$ & $(3.97 \pm 0.29)\,\text{E-7}$ & $0.016 \pm 0.001$ & $(9.15 \pm 0.59)\,\text{E-7}$ & $90.79 \pm 1.05$ \\
\bottomrule
\end{tabular}
}
\vspace{-4ex}
\end{table*}

\begin{table}[ht]
\centering
\vspace{.5ex}
\caption{Computational cost comparison}
\label{tab:cost}
\scriptsize
\renewcommand{\arraystretch}{1}
\resizebox{0.7\columnwidth}{!}{%
\begin{tabular}{lcc}
\hline
\textbf{Method} & \textbf{FLOPs (MFLOPs)} & \textbf{Wall Time (s)} \\
\hline
Proposed Method & $0.44$ & $5$ \\
MCMC Baseline & $106.10$ & $165$ \\
\hline
\end{tabular}
}
\vspace{-4ex}
\end{table}


\section{Discussion} \label{discussion}

Our proposed Structured Hierarchical Transformer was designed to infer $\nu$ oscillation parameters by leveraging attention to capture per-energy angular correlations and to learn how these signatures evolve across the energy spectrum.
From Table~\ref{tab:results_earth}, we note that our approach consistently produced RMSE values that are statistically comparable to the MCMC baseline.
Nevertheless, Table~\ref{tab:cost} shows that it does so at a dramatically lower computational cost, as it requires ${\sim}240$ times fewer FLOPs and is ${\sim}33$ times faster in average processing time.


Our method achieved statistically significantly lower RMSE values for $\theta_{12}$.
This result is noteworthy because oscillation maps have very low sensitivity to this parameter. 
Its variations induce relatively subtle and localized distortions in the oscillation patterns, which are easily masked by dominant effects from other parameters.
Our approach exploits these spatially localized signatures by preserving and learning from the full angular–energy structure of the probability maps.
In contrast, the MCMC baseline relies on a single aggregated discrepancy measure, the Frobenius norm between observed and simulated maps.
This global error can obscure small differences that manifest only in specific regions of the oscillation maps.
While longer MCMC chains could recover comparable performance, doing so would require an increased computational effort.

Conversely, we obtained significantly higher RMSE values when inferring $\delta_{\mathrm{CP}}$.
This is expected, as $\delta_{\mathrm{CP}}$ enters the oscillation probabilities only through subleading interference terms that induce subtle correlated variations across all appearance channels, not only on the three selected ones (Sec.~\ref{sec:data}).
Thus, residual information remains distributed over the full set of nine oscillation channels, which are instead jointly exploited by the MCMC approach.
Although incorporating additional channels or richer architectures could improve performance, the achieved accuracy is consistent with the intrinsic difficulty of constraining $\delta_{\mathrm{CP}}$, which is known to be unobservable in solar-neutrino measurements.
However, when high-precision estimates are required, our approach can serve as a fast front end to the MCMC pipeline. 
The parameters inferred by our model can be used as informed initial values for the MCMC sampler, which substantially accelerates convergence since far fewer simulator evaluations are needed compared to starting the chain from uninformed or random initial conditions.

Also from Table~\ref{tab:results_earth}, in the cases where we obtained comparable RMSE values with respect to the MCMC baseline, the Conformal DualAQD approach produced significantly narrower PIs while still being able to achieve an empirical (marginal) coverage of approximately $90\%$.
This highlights the method’s ability to generate adaptive intervals that respond more effectively to heteroskedastic uncertainty, concentrating width where the model is uncertain and tightening it where the predictions are locally reliable.
Crucially, the resulting interval widths are extremely small relative to the operating parameter ranges. 
For example, for $\theta_{12}$, the average PI width is about $0.13^{\circ}$, which is roughly two orders of magnitude smaller than its $4.6^{\circ}$ operating span.
Similar tight ratios are obtained for the rest of the parameters.
Therefore, these intervals provide sufficient resolution for $\nu$ oscillation analyses, as they restrict the plausible values to a very narrow portion of the physically allowed space, ensuring that the inferred parameters can be confidently localized within the global parameter landscape.

We further justify why conventional transformer architectures are not suitable for this task.
For instance, a 2-D attention model that attends over both $(E, \cos\theta)$ axes jointly could, in principle, learn the same mapping. 
However, it would be computationally heavier and would entangle angular and spectral processing from the start, losing the physically meaningful separation between per-energy angular patterns and inter-energy structure.
An alternative would be to flatten each oscillation map into a single long sequence and apply a standard transformer. 
This approach is substantially less efficient, as every attention layer would operate on a sequence whose length equals the total number of $(E, \cos\theta)$ bins, greatly increasing the parameter count and memory footprint. 

Vision Transformers mitigate this issue in image domains by dividing images into patches, embedding them, and operating attention over a reduced set of patch tokens rather than individual pixels. 
Such patch-based compression is unsuitable in our application, because small variations in the oscillation parameters induce subtle pixel-level deformations in the oscillation probability maps. 
As we verified in preliminary experiments, patch extraction
obscures these fine-grained variations, collapsing distinct oscillation patterns into nearly indistinguishable embeddings and ultimately degrading parameter identifiability.
By contrast, the proposed architecture preserves full $(E, \cos\theta)$ resolution while maintaining computational tractability. 

A limitation of the present approach lies in the choice of the reconstruction loss $\mathcal{L}_{\mathrm{rec}}$. 
While the Frobenius norm was sufficient to stabilize training and guide early optimization, future work should investigate loss functions tailored to capture the probabilistic structure of the maps and their inter-channel dependencies, such as KL divergence or Wasserstein distances.

Although our experiments are focused on atmospheric neutrinos, our framework is not tied to this $\nu$ source. 
The method relies only on the availability of simulated oscillation maps defined over energy and propagation variables, and therefore can be readily extended to other $\nu$ settings, such as accelerator-based experiments or long-baseline configurations, by adapting the simulator and the corresponding parameter ranges.

\section{Conclusion} \label{conclusion}

Neutrino oscillation parameter inference is a central task in neutrino physics, yet current workflows rely almost exclusively on MC methods, which are accurate but computationally expensive. 
These limitations become especially restrictive as next-generation telescopes, such as KM3NeT, yield higher-resolution data, which in turn demand greater precision and refined modeling. 
To our knowledge, no prior work attempted to infer oscillation parameters directly from oscillation probability maps, although these maps encode the full angular–energy structure that governs oscillation phenomenology.

Leveraging this structure, we proposed a framework based on Structured Hierarchical Transformers that process matter-effect oscillation maps and estimate their oscillation parameters. 
We also introduced Conformal DualAQD to produce 90\% prediction intervals alongside each parameter estimate, enabling uncertainty-aware inference.
Experimental results demonstrated that our model achieves accuracy comparable to the MCMC baseline while reducing computational cost by over two orders of magnitude. 
In addition, the obtained PIs remain narrow relative to each parameter’s operating range.
Thus, for any given set of maps, we can reliably localize their parameter values within a small region of the parameter space. 


This work represents an initial step toward map-based parameter inference on real observational data. 
Future work will model $\nu$ flux effects and develop a reconstruction stage that converts raw detector observations into oscillation probability maps. 
Once these steps are established, our models can process the reconstructed maps to recover the underlying oscillation parameters from actual observations, and to better model correlations and degeneracies among $\nu$ oscillation parameters.

\section*{Acknowledgements}

This research was funded by ANR (award ANR-19-CHIA-0017) and the Region Normandie (grant DEEPVISION).

\bibliographystyle{IEEEtran}
\bibliography{references}

\clearpage
\appendices

\section{Local Invertibility Analysis of the Neutrino Oscillation Simulator}~\label{app:A}

Parameter degeneracies (i.e., cases where different $\mathbf{p}$ produce almost indistinguishable $\mathbf{X}$) are known to occur in oscillation physics, such as in the cases of intrinsic, octant, and mass-ordering degeneracies~\cite{Minakata2001}. 
In addition, some parameters, such as $\delta_{\mathrm{CP}}$, are periodic, which can introduce global non-uniqueness; thus, the mapping $\mathbf{p} \mapsto \mathbf{X}$ is not globally invertible across the full physical domain.
However, within the restricted parameter ranges adopted in this study (Table~\ref{tab:param_ranges}), all parameters can be treated as effectively non-periodic.

To assess whether the simulator defines a locally one-to-one mapping between $\mathbf{p}$ and $\mathbf{X}$, we compute the numerical Jacobian of the forward model. Let $\mathbf{X}^f = F(\mathbf{p}) \in \mathbb{R}^D$ denote the flattened oscillation map produced by the simulator for a given parameter vector $\mathbf{p} \in \mathbb{R}^6$. The Jacobian matrix of $F$ is:
\[
J(\mathbf{p}) = \frac{\partial F(\mathbf{p})}{\partial \mathbf{p}} 
= 
\begin{bmatrix}
\frac{\partial X^f_1}{\partial p^{(1)}} & \cdots & \frac{\partial X^f_1}{\partial p^{(6)}} \\
\vdots & \ddots & \vdots \\
\frac{\partial X^f_D}{\partial p^{(1)}} & \cdots & \frac{\partial X^f_D}{\partial p^{(6)}}
\end{bmatrix}.
\]
Each column of $J(\mathbf{p})$ measures the infinitesimal change in the oscillation map with respect to one of the parameters. 
The matrix thus captures local parameter sensitivities and possible correlations among parameters.

Since the simulator is deterministic and differentiating it analytically is infeasible, we approximate $J(\mathbf{p})$ using central finite differences. 
For each parameter $p^{(i)}$, a small perturbation $\epsilon_i$ is applied while keeping the others fixed, giving  
\[
\frac{\partial F(\mathbf{p})}{\partial p^{(i)}} 
\approx 
\frac{F(\mathbf{p} + \epsilon^{(i)} \mathbf{e}^{(i)}) - F(\mathbf{p} - \epsilon^{(i)} \mathbf{e}^{(i)})}{2\epsilon^{(i)}},
\]
where $\mathbf{e}^{(i)} \in \{0,1\}^6$ is the $i$-th standard basis vector; i.e., $\mathbf{e}^{(i)}_k = 0$ $\forall k \neq i$ and $\mathbf{e}^{(i)}_i = 1$.
. 
Each resulting difference is assembled into the corresponding column of $J(\mathbf{p})$.  

We then perform a singular value decomposition (SVD) $J(\mathbf{p}) = U \Sigma V^\top,$
where $\Sigma = \mathrm{diag}(\sigma_1, \dots, \sigma_6)$ contains the singular values $\sigma_1 \ge \dots \ge \sigma_6 \ge 0$.  
According to the local inverse function theorem, the mapping $F$ is locally invertible around $\mathbf{p}$ if $J(\mathbf{p})$ has full column rank; i.e., if $\sigma_6 > 0$.  
The condition number $\kappa = \sigma_1 / \sigma_6$ provides a measure of local sensitivity.
That is, a high value indicates that small perturbations in $\mathbf{X}$ may lead to large variations in $\mathbf{p}$, whereas a moderate $\kappa$ suggests numerical stability.

In our finite-difference analysis using the matter-effect simulator, all singular values were found to be positive, with $\sigma_6 \approx 0.6$ and $\kappa \approx 130$ on average across 1,000 parameter realizations.
This $\kappa$ value represents a moderate level of anisotropy in the local parameter sensitivity. 
Although some parameter directions are relatively more sensitive than others, all six singular values remain well above zero, indicating that the Jacobian has full column rank and is locally invertible. 
The empirical diagnostics justify treating each oscillation parameter as independently inferable using decoupled scalar regressors $f^{(i)}$, while still allowing for consistent joint interpretation of the results.

\section{Surrogate Models}\label{app:B}

For the surrogate-based consistency step (Sec.~\ref{sec:training}), we require a differentiable approximation of the forward oscillation simulator $\mathcal{S}$.
Recall that $\mathbf{p} \in \mathbb{R}^6$ and $\mathbf{X}^{9 \times H \times W}$ denote the oscillation-parameter vector and the corresponding probability map produced by the full simulator, respectively.
Given that each transition channel $\nu_{\alpha} \to \nu_{\alpha}$ is conditionally dependent given $\mathbf{p}$, we construct a channel-wise surrogate simulator $\hat{\mathcal{S}} = \{ \hat{\mathcal{S}}^{(1)}, \dots, \hat{\mathcal{S}}^{(9)} \}$, where each component $\hat{\mathcal{S}}^{(c)}$ is a neural network trained to approximate the mapping from oscillation parameters to the probability values in channel $c$.

Formally, for each $c \in \{1, \dots, 9\}$, the surrogate model learns a function $\hat{\mathcal{S}}^{(c)}: \mathbb{R}^6 \mapsto \mathbb{R}^{H\times W}$ such that $\hat{\mathcal{S}}^{(c)} \approx \mathbf{X}^{(c)}$, where $\mathbf{X}^{(c)}$ denotes the $c$-th channel of the simulator output.
To reduce the learning complexity, we decompose the two-dimensional map $\mathbf{X}^{(c)}$ into individual pixels and learn a pixel-wise regression function.
Given a pixel location $(i,j)$ corresponding to a unique $(E, \cos\theta)$ pair, the surrogate input is the augmented vector $\mathbf{z}_{ij}^{(c)} = \big[ i, j, p^{(1)}, p^{(2)}, p^{(3)}, p^{(4)}, p^{(5)}, p^{(6)} \big] \in \mathbb{R}^8$ and the target is the oscillation probability value $\mathbf{X}_{ij}^{(c)}$.
Each $\hat{\mathcal{S}}^{(c)}$ is thus trained to approximate the pixel-wise function $g^{(c)}: \mathbb{R}^8\mapsto \mathbb{R}$, such that $g^{(c)}(\mathbf{z}_{ij}^{(c)}) \approx \mathbf{X}_{ij}^{(c)}$.
The full surrogate map $\hat{\mathcal{S}}^{(c)}(\mathbf{p})$ is obtained by evaluating $\hat{\mathbf{X}}_{ij}^{(c)}$ for all $i \in \{1, \dots, H\}, j \in \{1, \dots, W\}$.
Finally, the full surrogate simulator aggregates all channels: $\hat{\mathcal{S}}(\mathbf{p}) = \{ \hat{\mathcal{S}}^{(1)}(\mathbf{p}), \dots, \hat{\mathcal{S}}^{(9)}(\mathbf{p}) \}$.

Each surrogate model is a fully connected feed-forward network with five ReLU-activated hidden layers with 200, 500, 500, 100, and 50 units, respectively, followed by an output layer with a sinusoidal activation function.
Each surrogate model is a fully connected feed-forward network with five ReLU-activated hidden layers of sizes 200, 500, 500, 100, and 50, followed by an output layer with a sinusoidal activation. 
Similar to the setting explained in Sec.~\ref{sec:results}, we generated a dataset of 20,000 samples, using 80\% for training and 20\% for validation. 
The resulting RMSE between the target and predicted maps on the validation set yielded values of $(1.08, 7.26, 8.22, 1.87, 1.89, 2.66, 5.66, 1.89) \! \times \! 10^{-4}$ for the nine oscillation channels.
These small errors convey the surrogates' ability to produce high-fidelity oscillation maps. 

While additional architectural or training modifications could enforce the unitary constraints expected from oscillation probability maps, these simple models proved useful for guiding the training of our Structured Hierarchical Transformers, particularly during the early training stages. 
As noted in Sec.~\ref{sec:training}, our transformer models optimize a function $\mathcal{L}^{(i)}_{\mathrm{total}} = \mathcal{L}^{(i)}_{\mathrm{par}} + \lambda \mathcal{L}^{(i)}_{\mathrm{rec}}$, where the balancing weight $\lambda$ is scheduled to diminish over time. 
Thus, highly accurate surrogate models are not a priority toward the end of training, when parameter accuracy becomes the primary focus.

\begin{figure*}[!ht]
    \centering
    \includegraphics[width=0.85\linewidth]{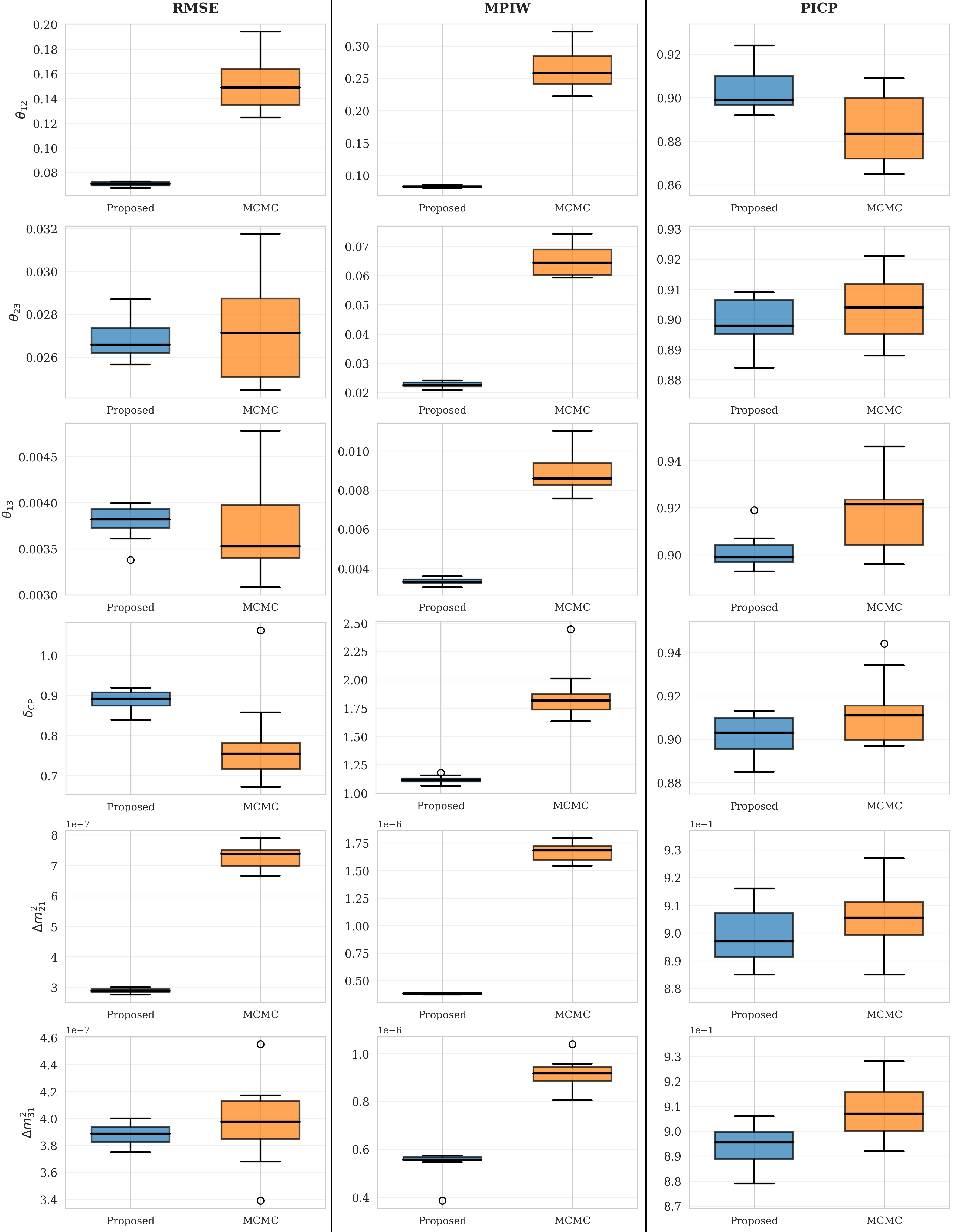}
    \caption{Box plots of the $\mathrm{RMSE}^{(i)}$, Rel. $\mathrm{RMSE}^{(i)}$, $\mathrm{MPIW}^{(i)}$, and $\mathrm{PICP}^{(i)}$ metrics obtained by both methods on the test sets.}
    \label{fig:box}
    \vspace{-2ex}
\end{figure*}

\section{Box Plots}\label{app:C}

Fig.~\ref{fig:box} depicts the distribution of the scores achieved by both methods during the test phase, which were summarized in Table II.
In these plots, the line through the center of each box indicates the median score, the edges of the boxes are the 25th and 75th percentiles, whiskers extend to the maximum and minimum points (not counting outliers), and outlier points are those past the end of the whiskers (i.e., those points greater than $1.5\times IQR$ plus the third quartile or less than $1.5\times IQR$ minus the first quartile, where $IQR$ is the inter-quartile range).


\end{document}